\documentstyle[11pt,newpasp,twoside,epsf]{article}
\reversemarginpar
\newcommand{\be}{\begin{equation}}
\newcommand{\ee}{\end{equation}}
\newcommand{\bea}{\begin{eqnarray}}
\newcommand{\eea}{\end{eqnarray}}

\begin{document}

\title{Scaling of ISM Turbulence: Implications for HI} 

\author{Jungyeon Cho, Alex Lazarian \& Huirong Yan}

\affil{ 475 N. Charter St., 
   Department of Astronomy, Univ. of Wisconsin, Madison, WI53706, USA;
   cho, lazarian, \& yan@astro.wisc.edu}


\begin{abstract}
Galactic HI is a gas that is coupled to magnetic
field because of its fractional ionization. 
Many properties of HI are affected by turbulence.
Recently, there has been a significant
breakthrough on the theory of magnetohydrodynamic (MHD) turbulence.
For the first time in the history of the subject we have
a scaling model that is supported by numerical simulations.
We review recent progress in studies of both incompressible
and compressible turbulence.
We also discuss the new regime of MHD turbulence that happens
below the scale at which conventional
turbulent motions get damped by viscosity.
The viscosity in the case of HI is being produced by neutrals
and truncates the turbulent cascade at parsec scales. We show
that below this scale magnetic fluctuations with a shallow spectrum
persist and point out to a possibility of the resumption of the
MHD cascade after ions and neutrals decouple.
We discuss the implications of the new insight into MHD turbulence
for cosmic ray transport and grain dynamics.

\end{abstract}

\section{Introduction}

 The interstellar medium (ISM) is clumpy and turbulent 
(Larson 1981; Myers 1983; Scalo 1987; see also Lazarian, Pogosyan \&
Esquivel, this volume, henceforth LPE02)
with an embedded magnetic field that influences almost all
of its properties. This turbulence 
that ranges from  AUs to
kpc (Armstrong et al.~1995, Stanimirovic \& Lazarian 2001, Deshpande et
al.~2000)
holds the key to many astrophysical
processes (e.g., star formation, fragmentation of molecular
clouds, heat and cosmic ray transport, magnetic reconnection).

All turbulent systems have one thing in common: they have large
 ``Reynolds number" ($Re\equiv LV/\nu$; L=characteristic
size of the system, V=velocity different over this size, and $\nu$=viscosity), 
the ratio of 
the time required for viscous
forces to slow it appreciably ($L^2/\nu$)
to
the eddy turnover time of a parcel of gas ($L/V$).
 A similar parameter, the ``magnetic
Reynolds number", $Rm$ ($\equiv LV/\eta$; $\eta$=magnetic diffusion), 
is the ratio of 
the magnetic field decay time ($L^2/\eta$)
to the eddy turnover time ($L/V$). 
The properties of the flows on all scales
depend on $Re$ and $Rm$. Flows with $Re<100$ are laminar; chaotic
structures develop gradually as $Re$ increases, and those with
$Re\sim10^3$ are appreciably less chaotic than those with
$Re\sim10^7$. Observed features such as star forming clouds are very
chaotic with $Re>10^8$ and $Rm>10^{16}$. 

{}Let us start by considering incompressible hydrodynamic turbulence,
which can be described by the 
Kolmogorov theory (Kolmogorov 1941).
Suppose that we excite fluid motions at a scale $L$.
We call this scale the {\it energy injection scale} or the
{\it largest energy containing eddy scale}. For instance, an obstacle
in a flow excites motions on the scale of the order of its size.
Then the energy injected at the scale $L$ cascades
to progressively
smaller and
smaller scales 
at a rate of eddies turning over, i.e. $\tau_l^{-1}\approx v_l/l$,
with the energy losses along the cascade being negligible\footnote{This
is easy to see as the motions at the scales of large eddies
have $Re\gg 1$ and therefore the energy loss
into heat is negligible over the eddy turnover time.}.
Ultimately, the energy reaches the molecular dissipation scale $l_d$,
i.e. the scale where the local $Re\sim 1$,
and is dissipated there.
The scales between $L$ and $l_d$ are called the {\it inertial range}
and it typically covers many decades. The motions over the inertial
range are {\it self-similar} and this provides tremendous advantages
for theoretical description. 

The beauty of the Kolmogorov theory that it does provide a simple
scaling for hydrodynamic motions. If
the velocity at a scale $l$ from the inertial range is $v_l$, the
Kolmogorov theory states that the kinetic energy ($\rho v_l^2\sim v_l^2$
as the density is constant) is
transferred to next scale within one eddy turnover
time ($l/v_l$). Thus within the Kolmogorov theory the energy
transfer rate ($v_l^2/(l/v_l)$) is scale-independent, and
we get the famous Kolmogorov scaling~~~~~$v_l \propto l^{1/3}$~~~. 

One-dimensional\footnote{Dealing with observational data, e.g. in
LPE02,
we deal with three dimensional energy spectrum $P(k)$, which, 
in isotropic turbulence, is
related to $E(k)$ in the following way: $E(k)=4\pi k^2 P(k)$.}
energy spectrum $E(k)$ is one of the most important 
quantities in turbulence theories.
Note that $E(k) dk$ is the amount of energy between the wavenumber $k$
and $k + dk$.
When $E(k)$ follows a power law, $kE(k)$ is the energy {\it near} the 
wavenumber $k\propto 1/l$.
Since $v_l^2$ represents a similar energy, $v_l^2 \approx kE(k)$.
Therefore, Kolmogorov scaling entails 
~~~~~~$E(k) \propto k^{-5/3}$~~~.

Kolmogorov scalings constitute probably the major advance of the microscopic
turbulence theory of incompressible fluids. They allowed numerous applications
in different branches of science (see Monin \& Yaglom
1975).  
However, astrophysical fluids are magnetized and the application of
the Kolmogorov scalings is not easy to justify. For instance,
dynamically important magnetic field should interfere with eddies
motions.    

Paradoxically, astrophysical measurements reveal 
the Kolmogorov spectra (see LPE02).
For instance, interstellar scintillation observations
indicate electron density spectrum follows a power law over
7 decades of length scales (see Armstrong et al.~1995).
The slope of the spectrum is very close to $-5/3$ for
$10^6 m$ - $10^{14} m$. At larger scales LPE02 summarizes the evidence of
$-5/3$ velocity power spectrum over pc-scales in HI.
Solar-wind observations provide {\it in-situ} measurements of the
power spectrum of magnetic fluctuations and 
Leamon et al. (1998) also obtained a slope of $\approx -5/3$.
Is this a coincidence?  
What properties the magnetized compressible ISM
is expected to have? These sort of questions we will deal with below. 

Here we describe our approach which is complementary to that in
Vazquez-Semadeni (this volume). The latter attempts to simulate
the ISM in its complexity by 
including many physical processes
(e.g. compression, self-gravity) simultaneously. 
As a downside of this, such simulations cannot distinguish between the 
consequences of different processes. We discuss a focused approach
when only after obtaining clear understanding on the simplest level,
the next level is attempted.
Therefore, we first consider incompressible MHD turbulence (\S2), then
discuss the viscous damping of incompressible turbulence in \S3, and
then we consider the effect of compression in \S4.
We discuss implications of our new understanding of MHD turbulence
for the problems of dust motion and cosmic ray scattering in \S5.

\section{Incompressible MHD Turbulence}

\subsection{Theory}

Observations suggest that the random component of magnetic field
is comparable with the regular component. 
Therefore, we consider the case that the rms velocity at the energy
injection scale is comparable to the Alfven speed of the mean field.
This means that we do not deal with the problem of magnetic
field generation, or magnetic dynamo (see Kulsrud \& Anderson 1992;
Vishniac \& Cho 2001; Maron \& Cowley 2001).

An ingenious model very similar in its beauty and simplicity 
to the Kolmogorov one has been proposed by 
Goldreich \& Sridhar (1995; hereinafter GS95) 
for incompressible MHD turbulence. Earlier theories
by Iroshnikov (1963) and Kraichnan (1965) did not account
for the anisotropy created by magnetic field. In GS95, however,
it is noted that the motions mixing magnetic field lines happen
essentially hydrodynamically and therefore 
the energy transfer rate is roughly
$\tau_{nl}^{-1} \approx k_{\bot}v_k$
where $k_{\bot}$ is the wavevector component perpendicular
to the local magnetic field. These mixing motions in the GS95
model are coupled with the wave-like motions parallel to
magnetic field which results in
{\it critical balance} condition~~~~$k_{\|} V_A \sim k_{\bot}v_k$~~~~,    
where $k_{\|}$ is the component of the wavevector parallel
to the local magnetic field. 

Conservation of energy in the turbulent cascade implies 
that the energy cascade rate $\dot{\epsilon}$ 
($= v_l^2/\tau_{nl}$) is constant:
\begin{equation}
\dot{\epsilon}\sim\frac{v_k^2}{\tau_{nl}}
                                          = \mbox{constant.}    \label{A6}
\end{equation}
Combining this with the critical balance condition
we obtain an anisotropy that increases with the decrease of the
scale~~~~~$k_{\|} \propto k_{\perp}^{2/3}$~~~~,
and a Kolmogorov-like spectrum for perpendicular motions
\begin{equation}
v_k \propto k_{\perp}^{-1/3}, \mbox{~~or,~} E(k) \propto k_{\perp}^{-5/3},
\end{equation}
which is not surprising as  magnetic field 
does not influence motions that do not bend it. At the
same time, the scale-dependent anisotropy reflects the fact that it
is more difficult for weaker small eddies to bend magnetic field.

GS95 shows the duality of turbulent motions. Those perpendicular to
magnetic field are essentially eddies, while those parallel to magnetic
field are waves. The critical balance condition couples the two types of
motions.

\subsection{Numerical simulations}
\begin{figure}[t!]
\vspace{-0.2in}
\begin{tabbing}
\epsfxsize=.5\columnwidth \epsfbox{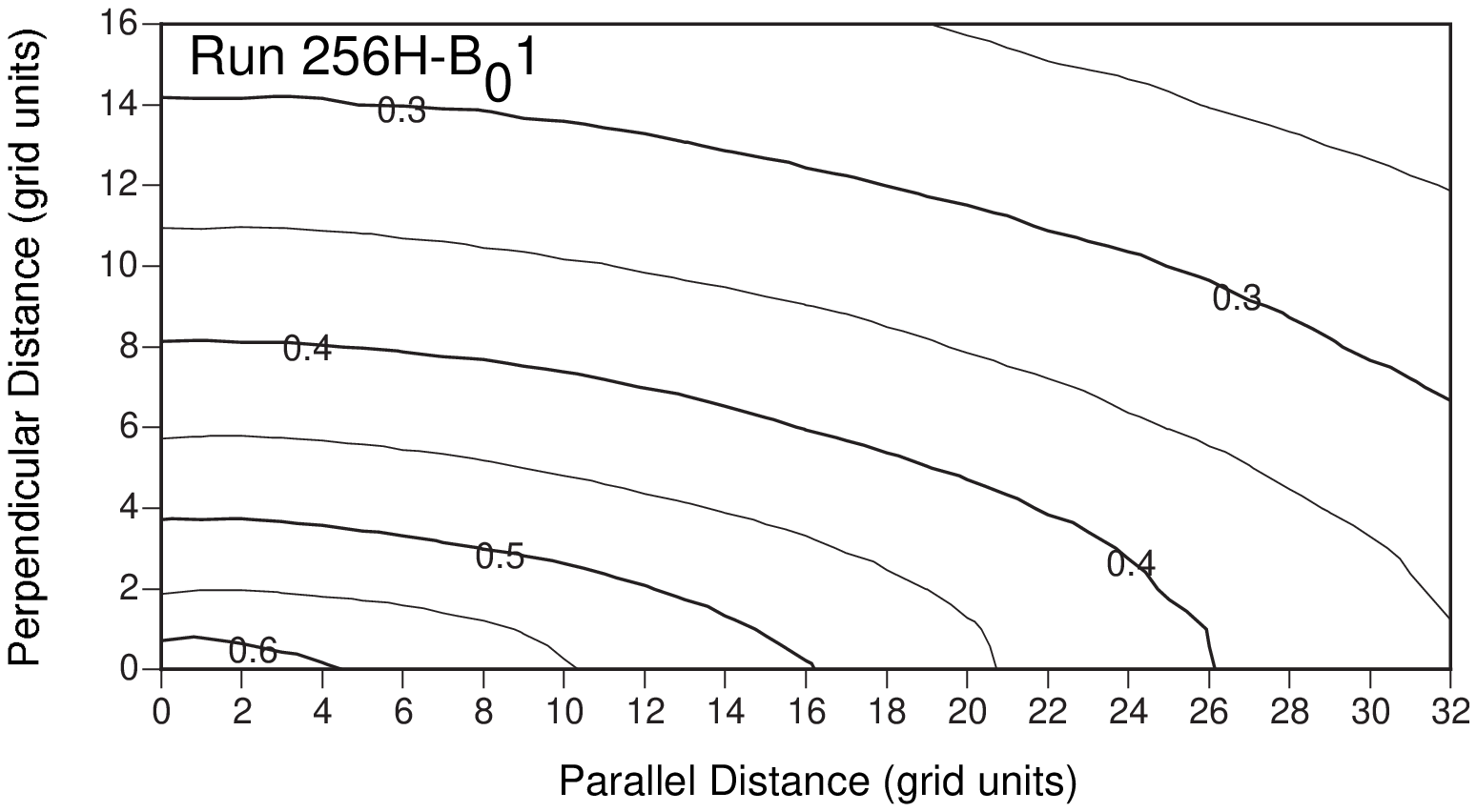}
\=
~~~~~\epsfxsize=.5\columnwidth \epsfbox{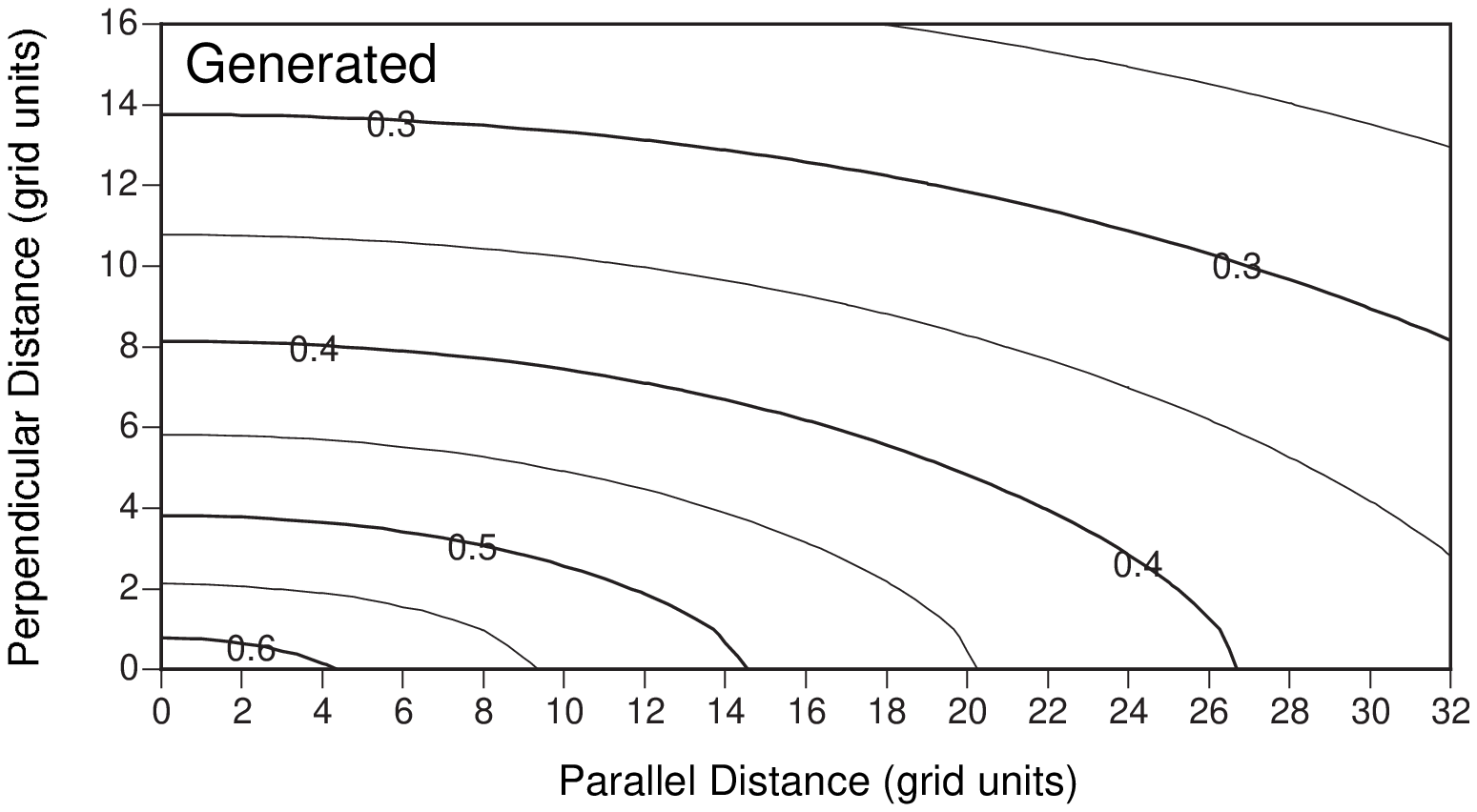} \\
~~~~~~~~~~~~~~~~~~~~~(a) \> ~~~~~~~~~~~~~~~~~~~~~~~~~~~(b) \\ 
~ \> ~ \\
~ \> ~ \\
~ \> ~ \\
~ \> ~ \\
\epsfxsize=.5\columnwidth \epsfbox[60 279 536 436]{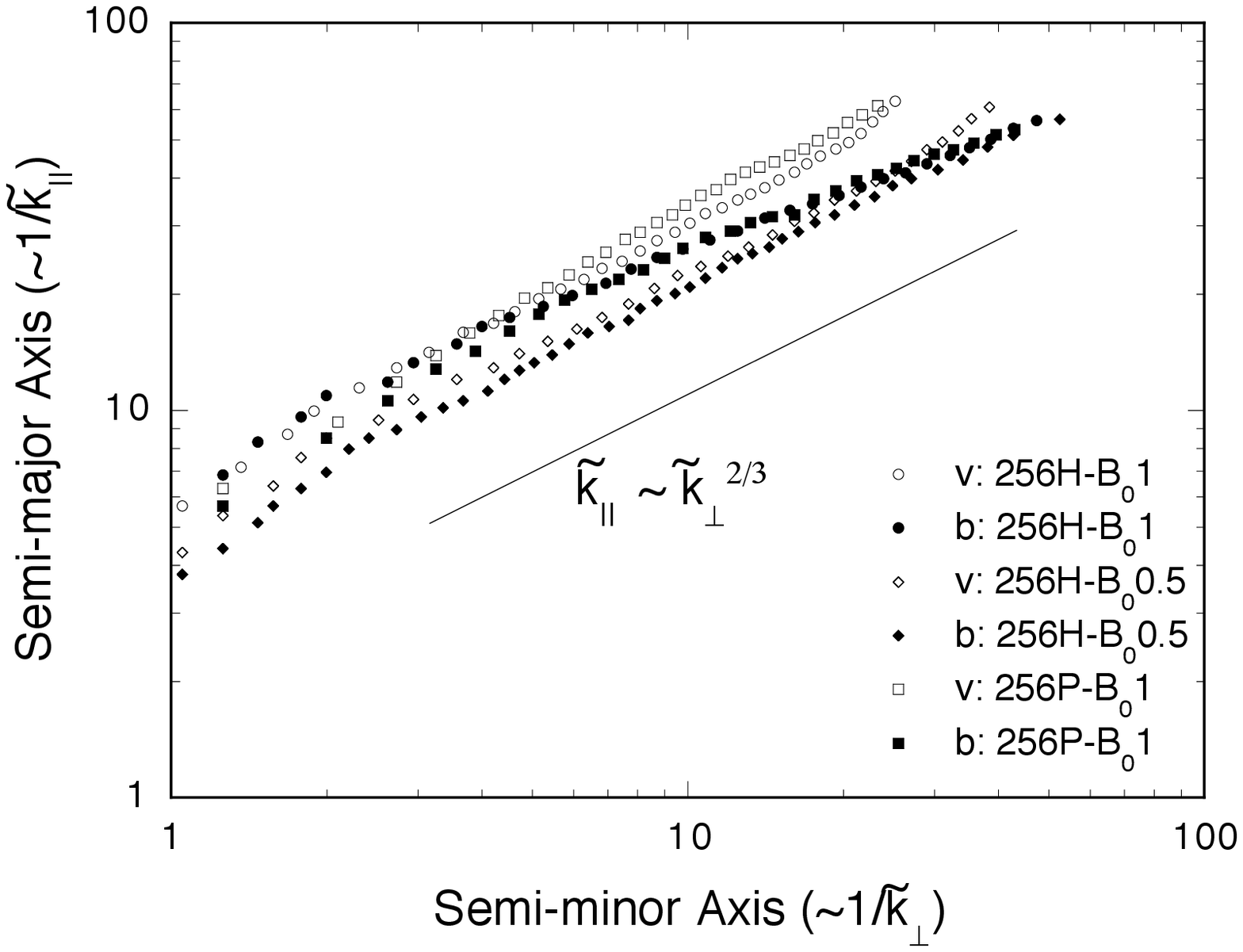}
\>
\epsfxsize=.5\columnwidth \epsfbox[20 460 516 620]{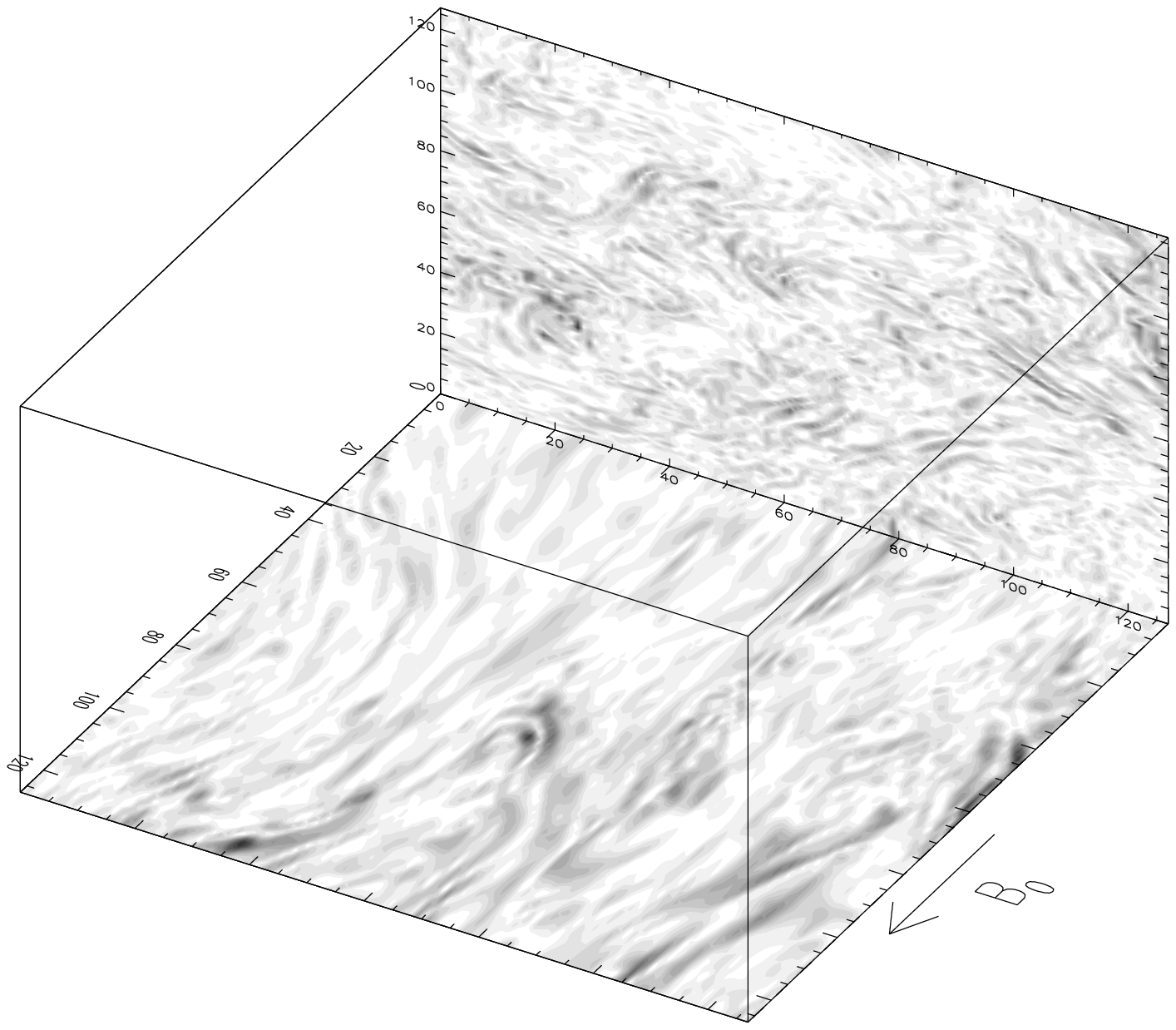}
\\
~ \> ~ \\
~ \> ~ \\
~ \> ~ \\
~ \> ~ \\
~~~~~~~~~~~~~~~~~~~~~(c) \> ~~~~~~~~~~~~~~~~~~~~~~~~~~~(d) \\
\end{tabbing}
\vspace{-0.3in}
\caption{(a) Velocity correlation function (VCF) from a simulation.
             Contours represent shape of different size eddies.
             The smaller contours (or, eddies) are
             more elongated.
         (b) VCF generated from equation (3).
         (c) Semi-major axis and semi-minor axis of contours of
             the same VCF (as in Fig.~1a).
             See Cho \& Vishniac (2000) for details.
         (d) Cross-sections of (a part of) the data cube
             Fourier components $k>20$. 
             Darker tones represent stronger magnetic field strength.
             The direction of the mean field is marked by an arrow.
             Magnetic fields show elongated structures along 
             the mean field ${\bf B}_0$ (bottom of the cube). 
             No systematic structures exist 
             in the perpendicular plane (back side of the cube). Figures by
             Cho, Lazarian, \& Vishniac.}
\end{figure}  

Numerical simulations by Cho \& Vishniac (2000) and 
Maron \& Goldreich (2001)
have confirmed GS95
theory, e.g. the Kolmogorov-type scaling\footnote{
Cho \& Vishniac (2000) obtained $E(k)\propto k^{-5/3}$ and Maron \& Goldreich
(2001) $E(k)\propto k^{-3/2}$.}
 and the
scale-dependent anisotropy  
($k_{\|}\propto k_{\perp}^{2/3}$),
and helped to extend it.
The important point that escaped earlier researchers was that
the 
scale dependent anisotropy can be measured only in a
local coordinate frame
 which is aligned with the locally averaged
magnetic field direction (see Cho \& Vishniac 2000).  
{}Further research in Cho, Lazarian,
\& Vishniac (2002a: hereinafter CLV02a) certified that in the local system of reference the
mixing motions perpendicular to magnetic field are identical to 
hydrodynamic motions (compare M\"{u}ller \& Biskamp 2000).

Fig.~1 illustrates some of our results.
 The contours of the equal correlation
obtained in 
Cho \& Vishniac (2000) are shown in
Fig.~1a and are 
 consistent with the predictions of the GS95 model.
{}Fig.~1c. shows that the semi-major axis ($1/k_{\|}$)
is proportional to the 2/3 power of the semi-minor axis ($1/k_{\perp}$),
implying that $k_{\|}\propto k_{\perp}^{2/3}$.
While one dimensional energy spectrum follows Kolmogorov spectrum,
$E(k)\propto k^{-5/3}$,
CLV02a showed that
3D energy spectrum is
\be
P(k_{\perp}, k_{\|})=(B_0/L^{1/3}) k_{\perp}^{-10/3}\exp\left(-L^{1/3}
     \frac{ k_{\|} }{ k_{\perp}^{2/3} } \right),
\label{tensor}
\ee
where $B_0$ is the strength of the mean field and $L$ is the scale of
the energy injection.
Velocity correlation from the 3D spectrum provides
an excellent fit to the numerical data (Fig. 1b). This allows practical
applications illustrated in \S5.

All in all, it has been shown that the magnetized incompressible turbulence
is both similar and dissimilar to the Kolmogorov turbulence. If
we consider motions perpendicular to magnetic field, they are 
hydrodynamic-like. This entails the Kolmogorov-type spectrum reported
by many observers. However, unlike Kolmogorov turbulence, the magnetized
one is anisotropic with the degree of anisotropy increasing with the
decrease of the scale. This entails much of a difference and requires
substantial revisions of many earlier calculations.

\subsection{Decay of turbulence}
Turbulence plays a critical role in molecular cloud support and star 
formation and the issue of the time scale of turbulent decay is vital for
understanding these processes.
If MHD turbulence decays quickly, then serious
problems face the researchers attempting to explain important observational 
facts, e.g.~turbulent  motions seen within molecular clouds without
star formation (Myers 1999) and rates of star formation (Mckee 1999).
Earlier studies attributed the rapid decay of turbulence to compressibility
effects (Mac Low 1999). GS95 predicts and numerical simulations, 
e.g.~CLV02a,  
confirm that turbulence decays rapidly even
in the incompressible limit. This can be understood if 
mixing motions perpendicular to magnetic field lines are considered. 
As we discussed earlier, such eddies, as in
hydrodynamic turbulence, decay in one eddy turnover time.

Below we consider the effect of 
imbalance on the turbulence decay time scale. Duality of the MHD turbulence
means that the
turbulence can be described by opposite-traveling wave packets.
`Imbalance' means that wave packets traveling in one
direction
have significantly larger amplitudes than those traveling in the other
direction.
In the ISM, many energy sources are localized both in space and time.
For example, in terms of energy injection, stellar outflows are essentially 
point energy sources.
With these localized energy sources, it is natural 
that interstellar turbulence be typically  imbalanced.

Here we show results of the
CLV02a study that demonstrates that imbalance does extend
the lifetime of MHD turbulence (see Fig.~2a).
We use a run on a grid of $144^3$ to investigate the decay time scale.
We run the simulation up to t=75 with non-zero driving forces.
Here $t$ is measured in the units of the large-scale eddy turnover time (L/V).
Then at t=75, we turn off the driving forces
and let the turbulence decay.
At t=75, the turbulence consists of upward (denoted as $+$) and downward moving
waves (denoted as $-$).
To adjust the degree of initial imbalance, we either increase or decrease 
the energy of the upward moving 
components and, by turning off the forcing terms, let the turbulence decay.
Note that the initial energy is normalized to 1.
The y-axis is the normalized total (=up $+$ down) energy.

\nopagebreak[3]
\begin{figure}[t!]
\plottwo{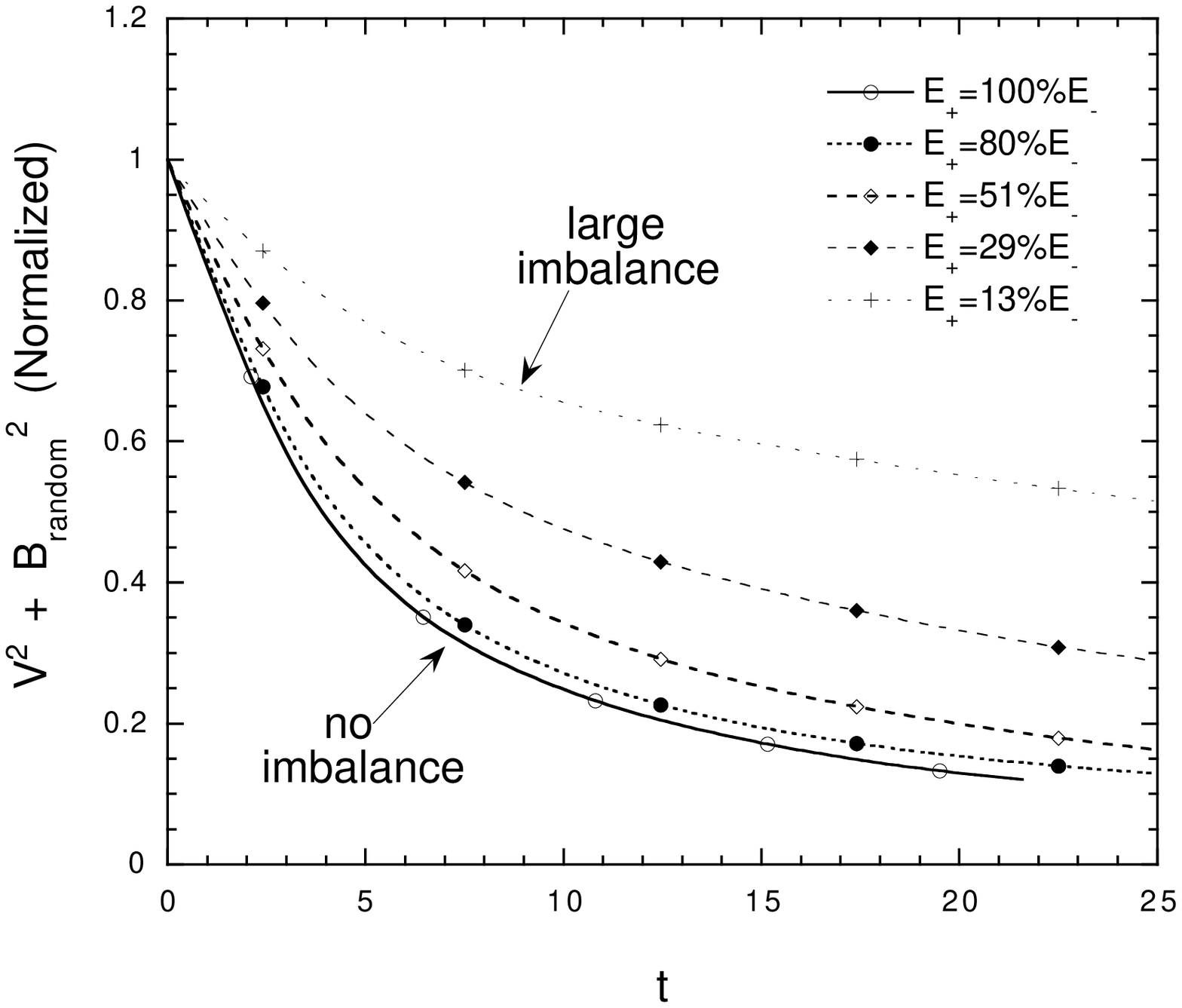}{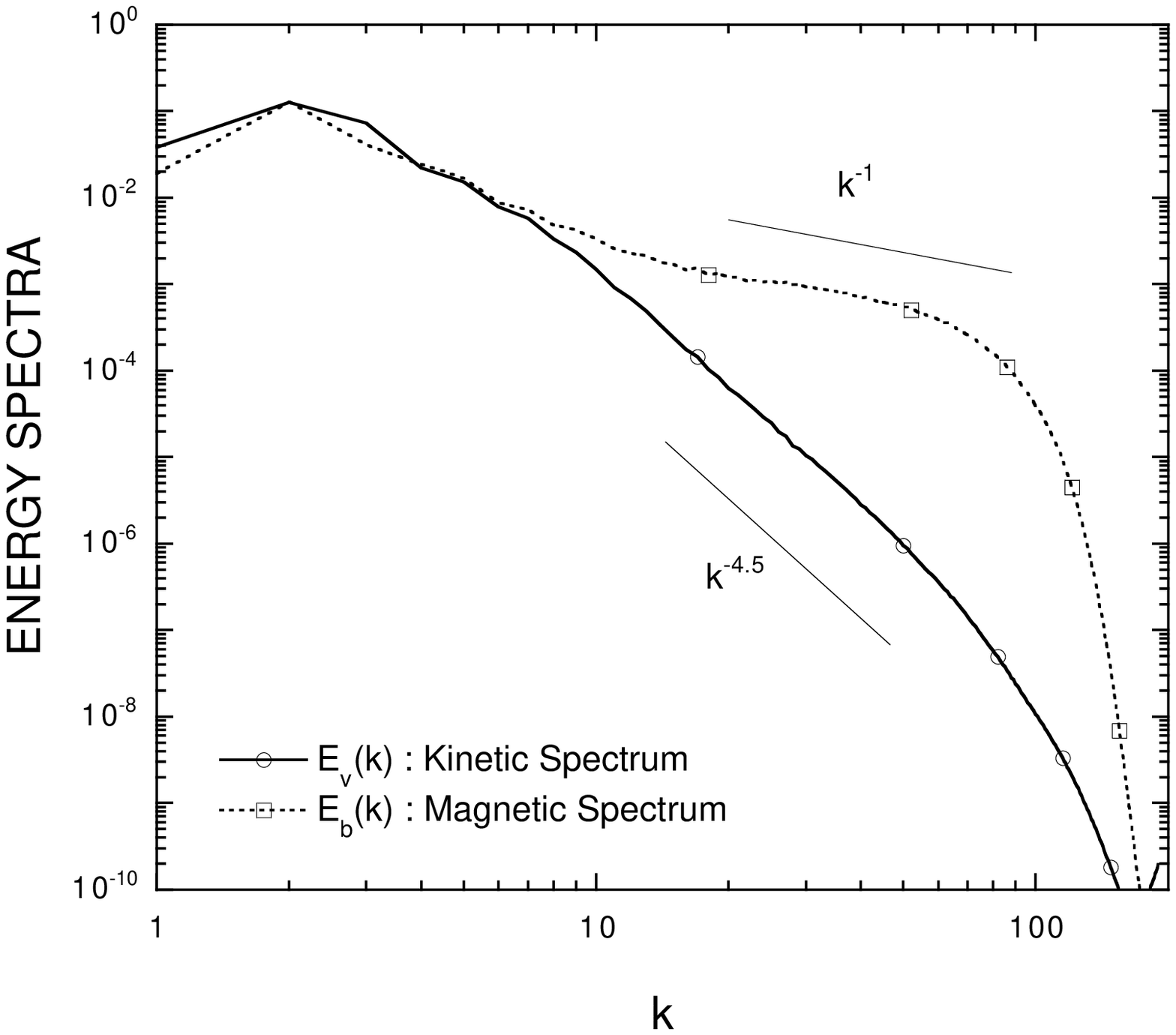}
\caption{(a) {\it (Left)} Imbalanced Decay. When imbalance is large,
             turbulence decays slow. From CLV02a.
         (b) {\it (Right)} Viscous damped regime. A new inertial range
             emerges below the viscous cut-off at $k\sim 7$.
             From Cho, Lazarian, \& Vishniac (2002b). 
             }
\vspace{-0.1in}
\end{figure}

The dependence of the turbulence decay time on the degree of imbalance
is an important finding. To what degree the results persist in 
the presence of compressibility is the subject of our current research.
It is obvious that our results are applicable to incompressible, namely,
Alfven motions. We show in \S4 that the Alfven motions
are essentially decoupled from compressible
modes. As the result we expect that the turbulence decay time may
be substantially longer than one eddy turnover time. 
  
\nopagebreak[3]
\section{Viscous Damped MHD Turbulence}
In hydrodynamic turbulence viscosity sets a minimal scale for
motion, with an exponential suppression of motion on smaller
scales.  Below the viscous cutoff the kinetic energy contained in a 
wavenumber band is 
dissipated at that scale, instead of being transferred to smaller scales.
This means the end of the hydrodynamic cascade, but in MHD turbulence
this is not the end of magnetic structure evolution.  For 
viscosity much larger than resistivity,
$\nu\gg\eta$, there will be a broad range of
scales where viscosity is important but resistivity is not.  On these
scales magnetic field structures will be created through a
combination of large scale shear and the small scale motions generated
by magnetic tension.  As a result, we expect
a power-law tail in the energy distribution, rather than an exponential
cutoff.  To our best knowledge, this is a completely new regime
of MHD turbulence.

In partially ionized gas, in general, and in HI, in particular, 
neutrals cause viscous damping on the scale of a fraction of
a parsec. The magnetic diffusion in those circumstances is
still negligible and pops in only at the much smaller scales $\sim 100km$. 
Thus exist a large
range of scales where physics is different from that in
the GS95 picture.

In Cho, Lazarian, \& Vishniac (2002b), we
numerically demonstrate the existence of the power-law
magnetic energy spectrum below the viscous damping scale.
We use a grid of $384^3$.
We use physical viscosity for velocity. The kinetic Reynolds number is
around 100. 
With this Reynolds number, viscous damping occurs around $k\sim 7$.
Here, $k\sim 7$ means that the wave length (or, the size of an eddy) 
is $\sim 1/7$ of a side of the computational box.
We use very small magnetic diffusion through the use of
hyper-diffusion of order 3. To test for possible
``bottle neck'' effects we also did simulations with normal 
magnetic diffusion and reproduced our results but with a reduced
dynamical range available.

In Fig.~2b, we plot energy spectra.
The spectra consist of several parts.
{}First, the peak of the spectra corresponds to the energy injection scale.
Second, for $2<k<7$, kinetic and magnetic spectra follow a similar slope.
This part is more or less a severely truncated inertial range for undamped MHD
turbulence.  
Third, the magnetic and kinetic spectra begin to decouple at $k\sim 
7$.
{}Fourth, after $k\sim20$, a new {\it damped-scale inertial 
range} emerges.
In the new inertial range, magnetic energy spectrum follows 
$E_b(k)\propto k^{-1}$, implying rich magnetic structures below the 
viscous damping scale.

We will present a theoretical model for this new regime and its
consequences for stochastic reconnection (Lazarian \& Vishniac 1999)
in an upcoming paper (Lazarian, Vishniac, \& Cho 2002). This model
implies that the ordinary MHD cascade resumes after the neutrals
and ions decouple. All the consequences of the new regime of the
MHD turbulence have not been appreciated yet, but we expect that it
will have a substantial impact on our understanding of the interstellar
physics.
In terms of H~I the result means
this small scale magnetic field might have some relation to 
the tiny-scale atomic structures (named by Heiles 1997), the mysterious
H~I absorbing structures on the scale from thousands to tens of
AU, discovered by Deiter, Welch, \& Romney (1976).

\nopagebreak[3]
\section{Compressible MHD Turbulence}
\begin{figure}[t!]
\vspace{-0.2in}
\begin{tabbing}
\epsfxsize=.475\columnwidth \epsfbox{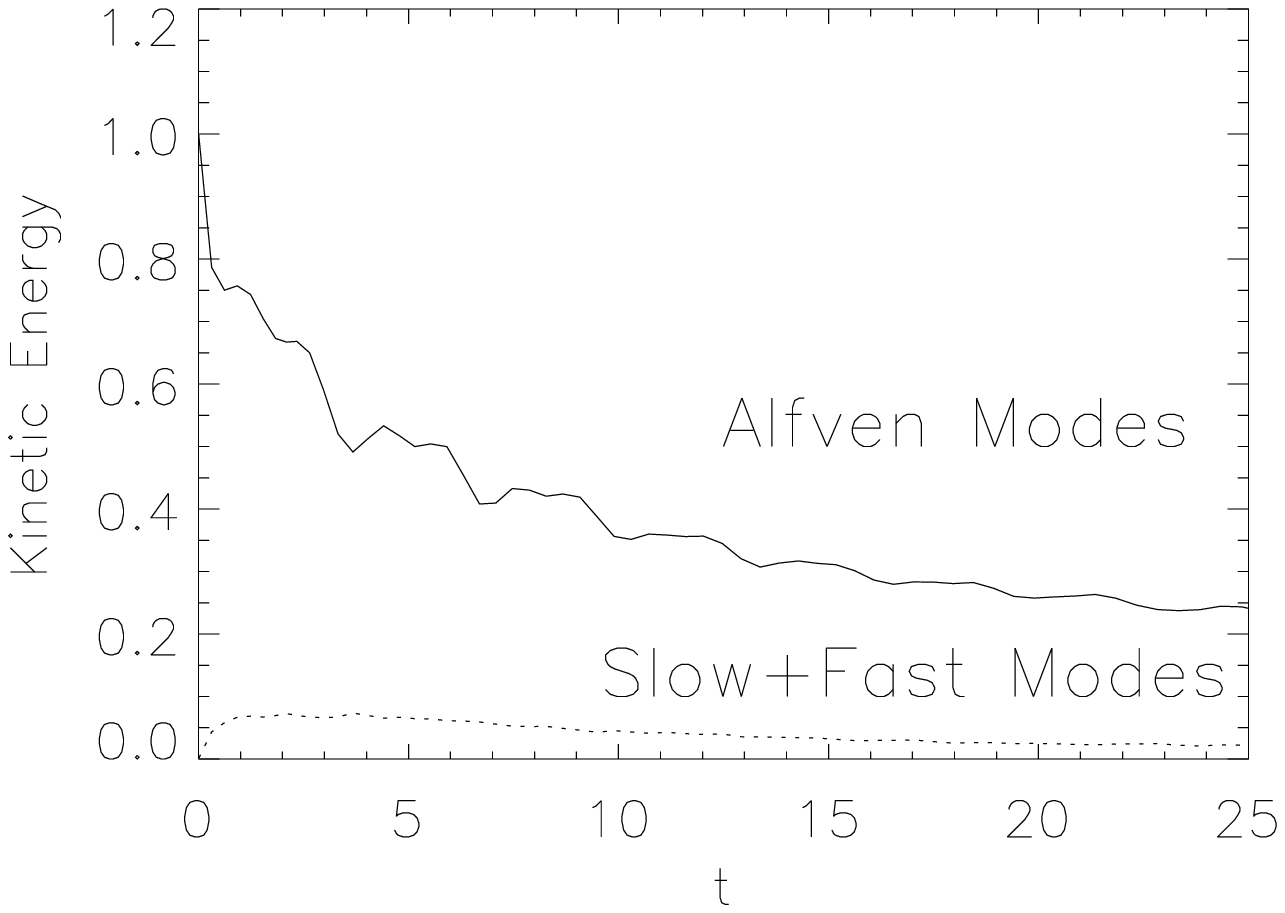}
\=
~~\epsfxsize=.5\columnwidth \epsfbox{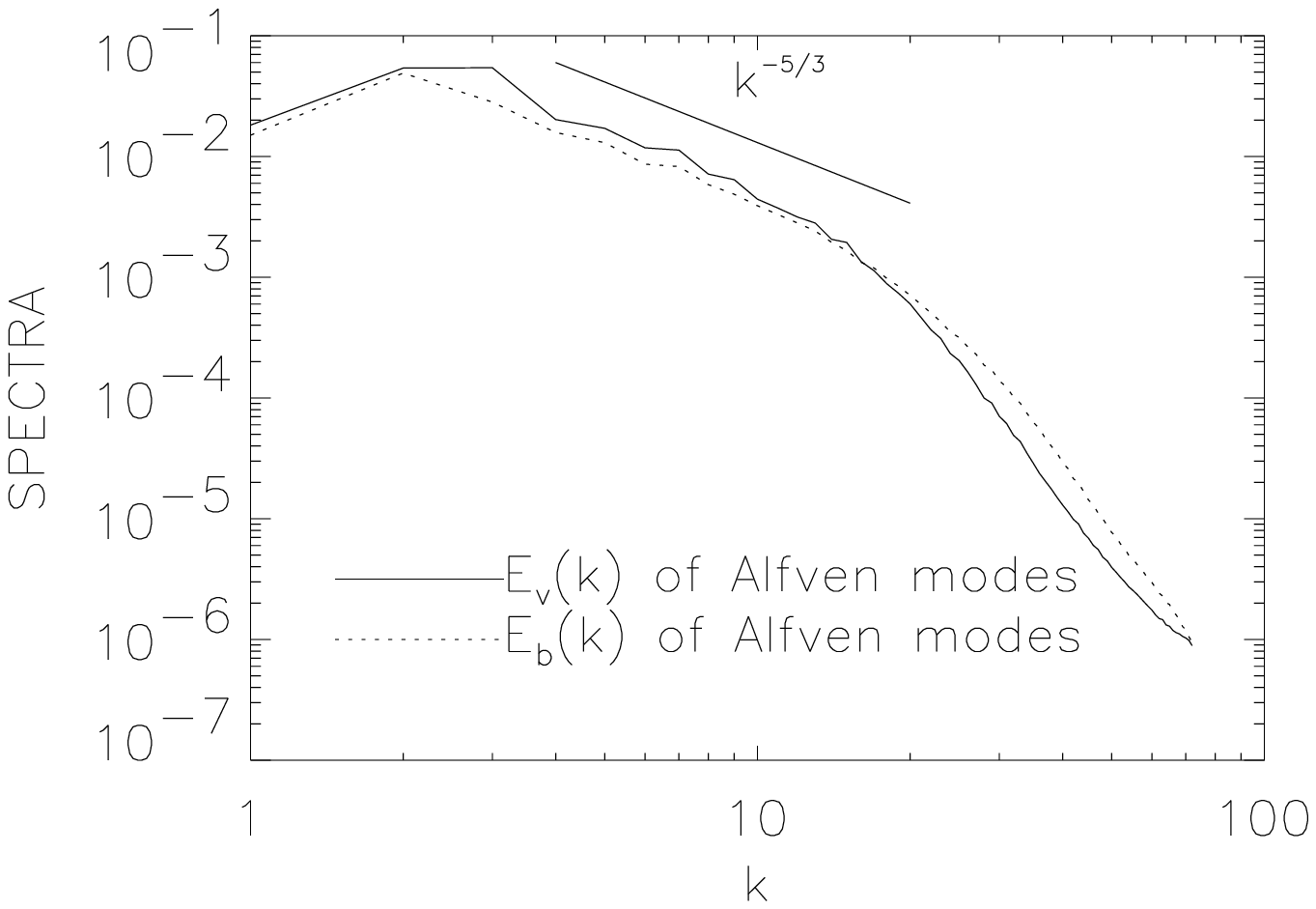} \\
~~~~~~~~~~~~~~~~~~~~~(a) \> ~~~~~~~~~~~~~~~~~~~~~~~~~~~(b) \\ 
~ \> ~ \\
~\epsfxsize=.46\columnwidth \epsfbox{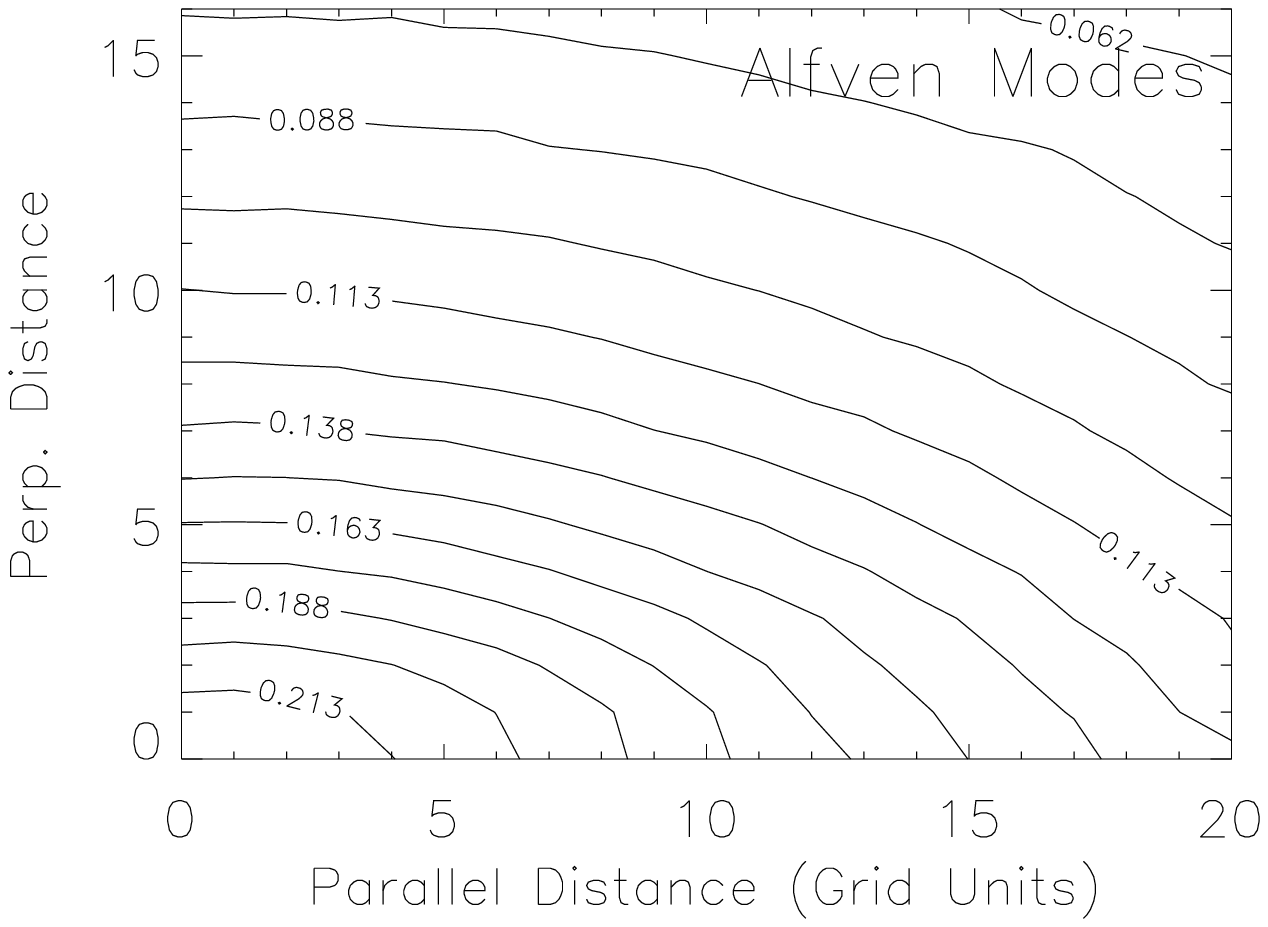}
\>
~~~~~~\epsfxsize=.46\columnwidth \epsfbox{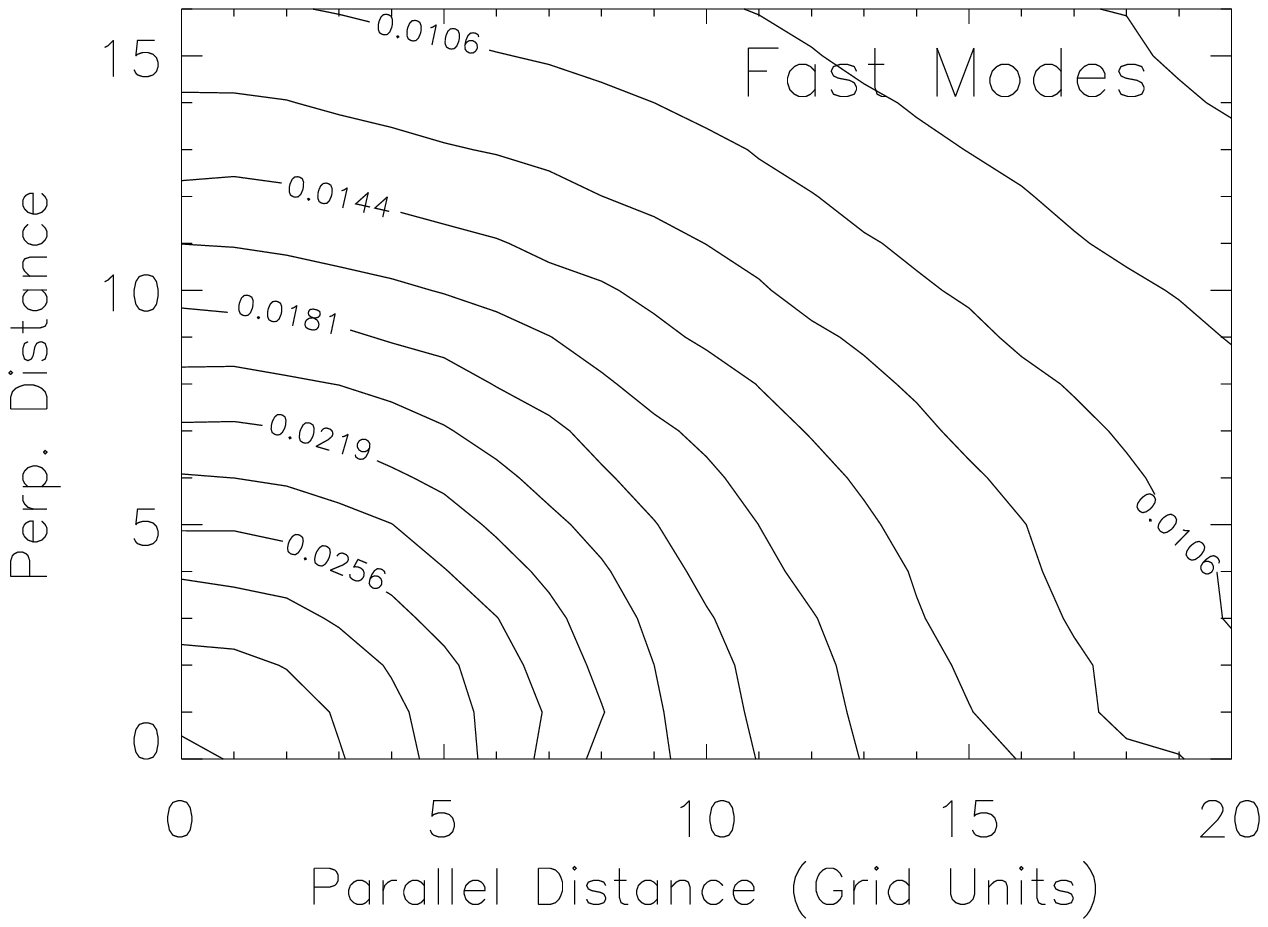}
\\
~~~~~~~~~~~~~~~~~~~~~(c) \> ~~~~~~~~~~~~~~~~~~~~~~~~~~~(d) \\
\end{tabbing}
\vspace{-0.3in}
\caption{Compressible MHD. (a) Alfven waves do not couple with other modes.
         (b) Alfven spectra follow a Kolmogorov-like power law.
         (c) Velocity correlation function (VCF) of 
             Alfven modes shows anisotropy similar to the GS95.
         (d) VCF of fast waves shows isotropy. From Cho \& Lazarian (2002). }
\vspace{-0.24in}
\end{figure}  

While the GS95 model describes incompressible
MHD turbulence well,
virtually no widely-accepted theory exists 
in compressible MHD turbulence regime
in spite of earlier theoretical attempts (e.g., Higdon 1984).
Again valuable insights are available from
numerical simulations focused on obtaining microphysical picture.
In this section, we will show our numerical results (Cho \& Lazarian 2002,
in preparation).

In compressible regime, there are 3 different MHD modes - Alfven, slow, and
{}fast modes. MHD turbulence can be viewed as interactions of these modes.
Therefore, it is important to study coupling between different modes.

Alfven modes are the least susceptible modes to damping mechanisms (see
Minter \& Spangler 1997).
Therefore, we mainly consider the transfer of energy from  
Alfven waves to compressible MHD waves (i.e.
slow and fast modes).
We carry out a simulation 
to check the strength of the coupling.
We first obtain a data cube from a incompressible numerical simulation.
Then we retain only Alfven modes and, using a compressible MHD code, 
let the turbulence decay.
{}Fig.~3a shows time evolution of kinetic energy.
Solid line represents kinetic energy of Alfven  modes.
It is clear that Alfven waves do not generate slow and fast modes
efficiently.
This means that Alfven modes poorly coupled with slow and fast modes.
Therefore, we expect that Alfven modes follow the same scaling
relation as in incompressible case.
Indeed, Fig.~3b shows that Alfven energy spectra 
follow a Kolmogorov-like spectrum.
The slow modes follow a bit steeper spectrum.

Fig. 3c shows that the anisotropy of Alfven waves, is
compatible with the GS95 model.
However, compressible modes do not necessarily show different behavior.
 For instance, Fig.~3d indicates that
velocity of fast modes is almost isotropic.




\section{Applications}

How properties of turbulence change with the scale is extremely important
to know for many astrophysical problems. Therefore we expect a wide range
of applications of the established scaling relations. Here we show how
recent breakthrough in understanding of MHD turbulence affects a few
selected issues. 

\subsection{Cosmic ray propagation}
The propagation of cosmic rays is mainly determined by their interactions
with the electromagnetic fluctuations in interstellar medium. 
The resonant interaction of cosmic ray particles
with MHD turbulence has been suggested by many authors as the main
mechanism to scatter and isotropicize cosmic rays. The
turbulence, that is normally considered is the  
{\it isotropic} turbulence with the Kolmogorov spectrum (see  
Schlickeiser \& Miller 1998). We know that this is not
a valid model. How should these calculations be modified?

The essence of the mechanism is rather simple.
Particles moving with velocity $v$ interact with resonant Alfven wave
of frequency \( \omega =k_{\parallel }v\mu +n\Omega  \) (\( n=\pm 1,2... \)),
where \( \Omega =\Omega _{0}/\gamma  \)
is the gyrofrequency of relativistic particles, \( \mu  \) is the
cosine of the pitch angle. From the resonant
condition above, we know that the most important interaction occurs at 
\( k_{\parallel }\sim \Omega /v\mu \sim (\mu r_{L})^{-1} \),
where \( r_{L} \) is Larmor radius of the high-energy particles. 

Adopting quasi-linear theory, we (Yan \& Lazarian 2002a)
calculated the scattering efficiency of both 
isotropic and anisotropic turbulence. The results are shown in Fig.~4a.
We see that the scattering is substantially suppressed, compared to the 
Kolmogorov turbulence. This happens, first of all, 
because most turbulent energy in GS95 turbulence goes to
\( k_{\perp } \) so that there is much
less energy left in the resonance point \( k_{\parallel }=(\mu r_{L})^{-1} \).
Furthermore, \( k_{\perp }\gg k_{\parallel } \) means \( k_{\perp }\gg r_{L}^{-1} \)
so that cosmic ray particles surf lots of eddies during one gyration. 
The random walk decreases the scattering efficiency by a factor of 
\( (\Omega /k_{\perp }v_{\perp })^{1\over 2}=(r_{L}/l_{\perp })^{1\over 2} \),
where \( l_{\perp } \) is the turbulence scale perpendicular to magnetic
field.

Thus the gyroresonance is not an effective way to provide scattering 
of cosmic rays if turbulence is injected on the large scales.
Indeed, we discussed earlier that the turbulent eddies get more
and more elongated as the energy cascades to smaller scales.
However, if the turbulence energy is injected isotropically
at small scales, 
the turbulence should be 
more isotropic and the scattering will be more efficient.

There is another property of turbulence that escaped the attention
of earlier researchers. It is known that 
when cosmic rays stream at a velocity much larger than Alfven velocity,
they can excite resonant MHD waves which in turn scatter cosmic rays.
This is so called streaming instability. 
It is usually assumed that the instability
can provide the confinement for cosmic rays 
with energy less than 100GeV (Cesarsky
1980). However, this is true only in an idealized situation when 
there is no background MHD turbulence. 
We discussed earlier that the rates of turbulent
decay are very fast and therefore the excited perturbations should
vanish quickly. In Yan \& Lazarian (2002a) we find that
the streaming instability is only applicable to particles with energy
\( <0.15GeV, \) which is less than the energy of most cosmic ray
particles. So we don't think self-confinement mechanism works as discussed
by previous authors. 

Another more efficient resonant process is the 
transit-time damping (TTD) invoking
the  fast MHD mode. However, in a hot plasma this mode
gets Landau-damped before nonlinear cascading transfers the energy
to the small scales. 
The slow rate of turbulent energy transfer for fast waves
follows from the marginal coupling of fast waves that we discussed in \S4.
Thus the range of applicability of the TTD is
limited.

All these findings bring more support to the alternative
picture of cosmic ray diffusion advocated by Jokipii 
(see Kota \& Jokipii 2000). In this picture the cosmic rays
follow magnetic field lines, but the field is wandering.
The rate of this wandering can be calculated from the established turbulence
scaling.

\begin{figure} [t!]
{\centering \leavevmode
\epsfxsize=.45\columnwidth \epsfysize=.45\columnwidth \epsfbox{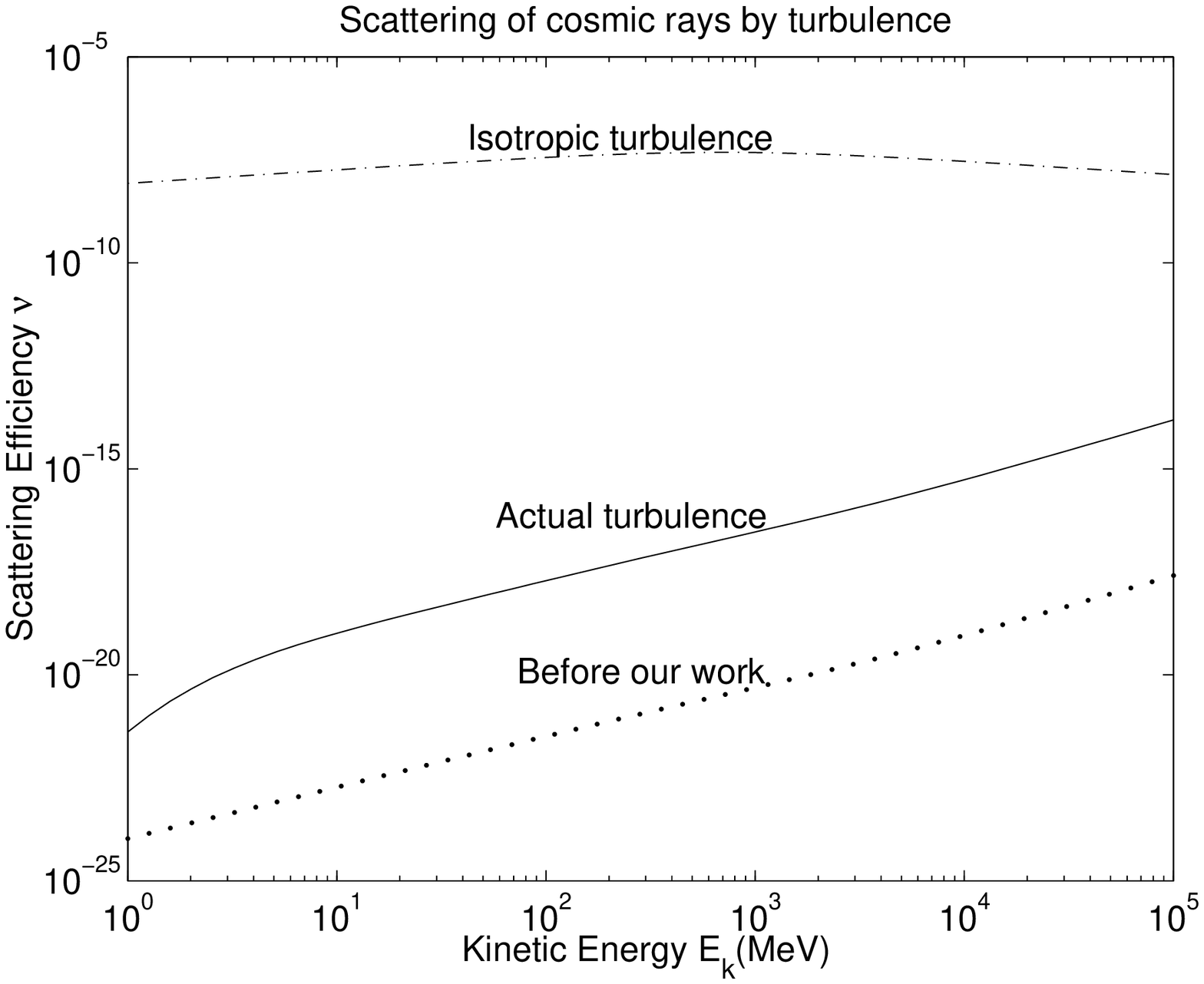} \hfil
\epsfxsize=.45\columnwidth \epsfysize=.45\columnwidth \epsfbox{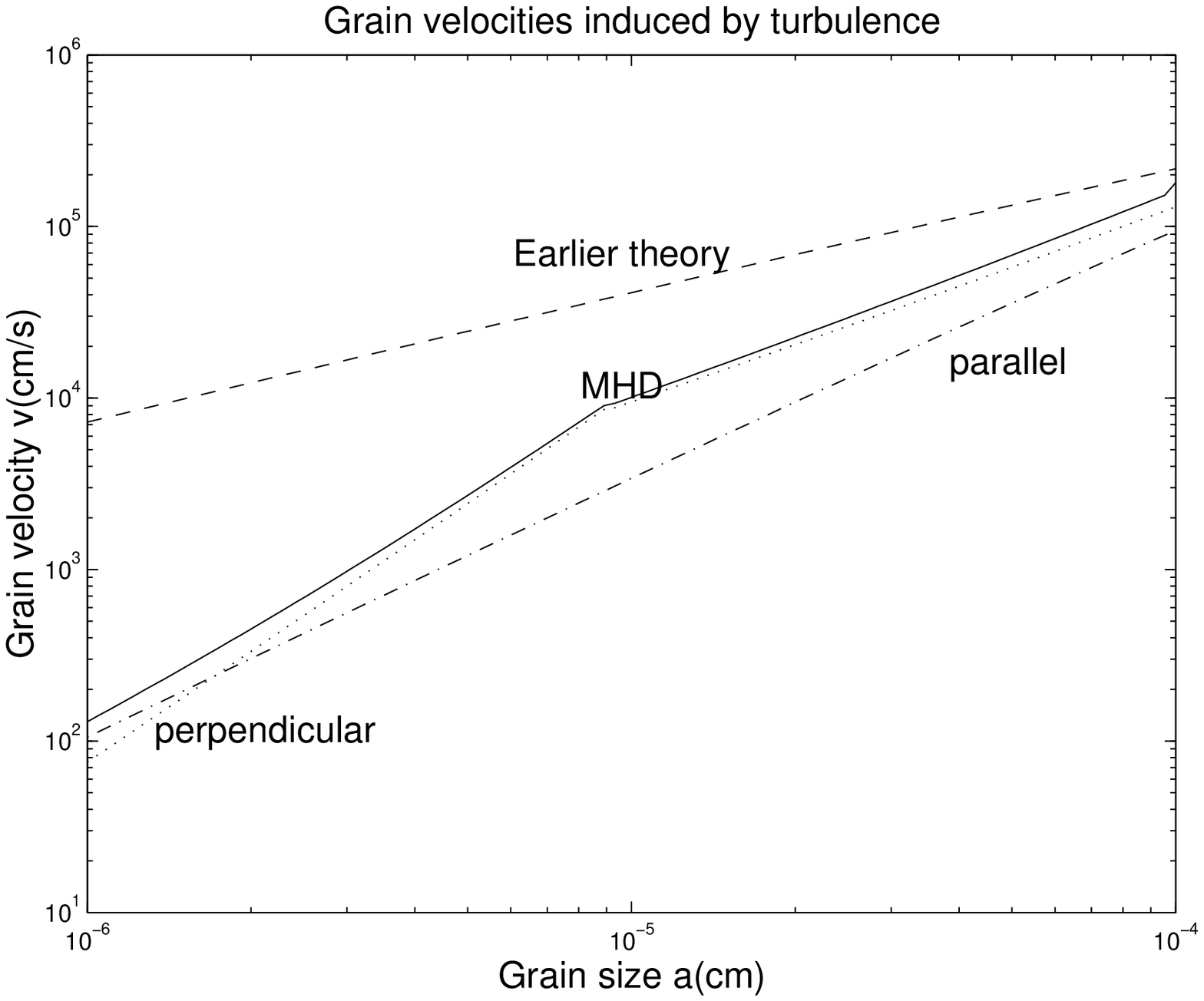} 
}
 \caption{
Applications. (By Lazarian \& Yan).
{\it Left:} Cosmic ray scattering for the actual
MHD turbulence is reduced substantially compared to scattering by
isotropic turbulence but larger than estimates in Chandran (2000). 
{\it Right:} The dust acceleration by turbulence is reduced 
compared to the accepted estimates in Draine (1985).}
\end{figure}

\subsection{Grain dynamics}
Turbulence induces dust grain relative motions and causes 
grain-grain collisions. Those
determine grain size distribution which affects most of the dust properties
including absorption and  H$_2$ formation. Unfortunately, as in the
case of cosmic rays, the earlier 
research appealed
to hydrodynamic turbulence to predict grain relative velocities (see Kusaka et
al.~1970; Volk et al.~1980; Draine 1985; Ossenkopf 1993; Weidenschilling
\& Ruzmaikina 1994). 

The differences between the hydrodynamic and MHD calculations stem from
(a) grain being charged and thus coupled with magnetic field, (b) anisotropy
of MHD cascade, and (c) direct interaction of charged grains with
magnetic perturbations. Effects (a) and (b) are considered
in Lazarian \& Yan (2002), while (c) is considered in Yan \& Lazarian
(2002b; in preparation). As the consequence of these studies the picture of the
grain dynamics is substantially altered.
 
Consider grain charge first.
If grain's Larmor time \( \tau _{L}=2\pi m_{gr}c/qB \)
is shorter than gas drag time \( t_{drag} \), grain perpendicular
motions are constrained by magnetic field. Their velocity dispersion are 
determined by the turbulence eddy whose 
turnover period is \( \sim \tau _{L} \) instead of drag time (Draine 1985). 

Accounting for the anisotropy of MHD turbulence it is convenient to consider 
separately
grain motions parallel and perpendicular to magnetic field. The 
perpendicular motion is 
influenced by the Alfven mode, 
which has a Kolmogorov spectrum. The parallel motion is subjected to 
compressible modes 
with scaling \( v_\parallel\propto k_\parallel^{-1/2} \).
In addition we should account for turbulence damping via viscous forces. 
When the eddy turnover time 
is of the order of \( t_{damp}\sim \nu _{n}^{-1}k_{\perp }^{-2} \), 
the turbulence is viscously damped. 
Thus grains sample only a part of the eddy before gaining the 
velocity of ambient gas if \( \tau_{L} \) or \( t_{drag} < t_{damp} \). 
The results are shown in Fig.~4b.

The direct interaction of the charged grains with turbulent magnetic field
results in a stochastic acceleration that can potentially
provide grains with supersonic
velocities.

\section{Summary}
Recently there have been significant advances in the field of
MHD turbulence:
 
1. The first self-consistent model (GS95) of incompressible MHD turbulence
that is supported by both numerical simulations and observations has 
been suggested.
The major predictions of the model are scale-dependent anisotropy 
($k_{\|}\propto k_{\perp}^{2/3}$) and
a Kolmogorov energy spectrum ($E(k)\propto k^{-5/3}$).

2. Simulations of compressible MHD turbulence show that
there is a weak coupling between Alfven waves and compressible MHD waves and
that the Alfven modes follow the Goldreich-Sridhar scaling.
{}Fast modes, however, decouple and exhibit isotropy.

3. On the contrary to the general belief, in typical interstellar
conditions,
magnetic fields can have rich structures below the scale
at which motions are damped by viscosity created by neutrals
(ambipolar diffusion damping scale).

These advances change a lot in our understanding of many
fundamental interstellar processes, e.g. 
cosmic-ray propagation and grain dynamics. In terms of HI
they show a way to account for the formation of structures
at very small scales. More discoveries are surely to come!


\acknowledgments{
   We thank Ethan T. Vishniac and Peter Goldreich for valuable discussions.
We acknowledge the support of NSF Grant AST-0125544.
This work was partially supported by National Computational Science
Alliance under CTS980010N and AST000010N and
utilized the NCSA SGI/CRAY Origin2000.
}
\nopagebreak

\end{document}